# Density of states and ground state magnetic ordering of the triangular lattice three-state Potts model

M.A. Magomedov [1, 2, *], A.K. Murtazaev [1]

[1] Institute of Physics, Dagestan Scientific Center, Russian Academy of Sciences,
367003, Makhachkala, M. Yaragskogo st., 94, Russia.

[2] Dagestan Scientific Center, Russian Academy of Sciences,
367025, Makhachkala, M. Gadjieva st., 45, Russia.

[*] e-mail: magomedov_ma@mail.ru

**Abstract.** This study present a Monte Carlo investigations of low-temperature magnetic ordering and phase transitions in three-state Potts model on triangular lattice with various exchange interactions between nearest ($J_1$) and next-nearest ($J_2$) neighbors. The density of states for varying $J_1$ and $J_2$ are calculated. The magnetic structure of the ground state for various $J_1$ and $J_2$ are obtained. The critical temperature are calculated and the order of the phase transition determined. The density of states difference (DOSD), $\ln g(E+\Delta E) - \ln g(E)$, and histogram analysis methods are used to investigate the order of the phase transitions. The frustrated regions are determined. It is shown, that for negative $J_1$ the high degeneration of the ground state are in fully frustrated area $-1 \leq J_2/|J_1| \leq -0.2$. For positive $J_1$ frustration are occurred in area $-1 \leq J_2/J_1 \leq -0.5$, but only in point $J_2/J_1 = -1$ the system have a high degeneration and are fully frustrated. The phase diagram of the three-state triangular Potts model are show.

## 1. Introduction

Investigations of the low-temperature ordering, phase transition and frustration phenomena in low-dimensional systems have a great scientific interest [1, 2]. In this paper, we investigate the three state Potts model on the triangular lattice using the Wang-Landau algorithm of Monte Carlo method. We analyze the ground state ordering and phase transitions in in this model varying the interaction between sites. This model applicable to many physical systems, such as delafossite-type compounds, layered magnets, superconducting films, simple fluids, polycrystalline materials and others [1-8].

The delafossite-type compounds with the general formula $ABO_2$ ($CuFeO_2$, $PtCoO_2$, $PdCoO_2$, $PdRhO_2$, $PdCrO_2$) have attracted much interest due to the quasi two-dimensionality of the lattice and the triangular arrangement of the transition-metal atoms. Characterized by a wide range of possible compositions, the delafossite oxides exhibit a significant richness in properties, enabling diverse technical applications, e.g. as a catalyst, in p-type conduction oxides, as a cathode in Li-ion batteries, or as luminescent materials. Furthermore, delafossite has special magnetic, photo- and electrochemical and antiviral properties. A strong anisotropy of those structural and physical properties makes this family of materials highly interesting [5-8].

**2. Model and Method**

The Hamiltonian of the $q=3$ Potts model with interaction of nearest and second nearest neighbors sites can be written as [9, 10]

$$H = -J_1 \sum_{\langle i,j \rangle} \cos\theta_{i,j} - J_2 \sum_{\langle i,k \rangle} \cos\theta_{i,k}, \qquad (1)$$

where the first sum corresponds to the exchange interactions between the nearest-neighbor spins characterized by the constant $J_1$, the second term describes the interaction between the next-nearest-neighbors with the interaction constant $J_2$; $\theta_{i,j}$, $\theta_{i,k}$ are the angles between the interacting spins $S_i$–$S_j$ and $S_i$–$S_k$, respectively, and $\theta_{i,j}$ takes on three values of 0°, 120°, 240°. Fig. 1. Show the description of the three-state Potts model with interaction between the nearest-neighbor and next-nearest-neighbors spins. Inset of figure show the color representation of the spin orientations.

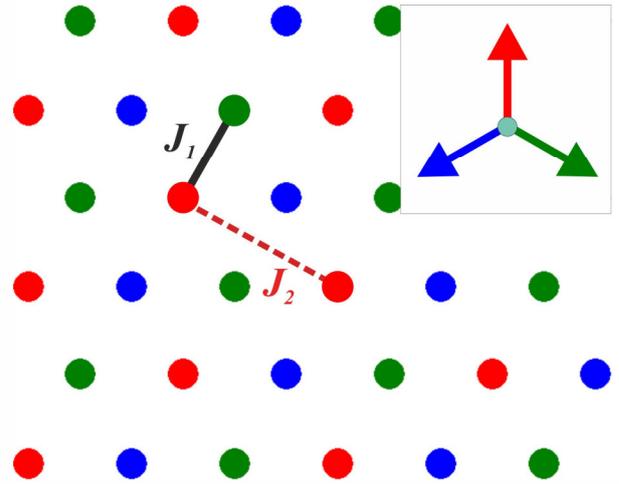

Fig. 1. The three-state Potts model on triangular lattice (inset show the color representation of the spin orientations).

If two neighboring spins take the same state, then the energy coupling is $-J_1$ for nearest neighbors and $-J_2$ for second nearest, otherwise it becomes $+0.5J_1$ and $+0.5J_2$. In this work the value $J_1$ fixed $J_1 = 1$ or $J_1 = -1$ and $J_2$ vary from -2 to 0.

Monte Carlo simulation is a standard method used in many field of physics, including the study of phase transition and critical phenomena. We use Wang-Landau algorithm to calculate the density of states of the models. This algorithm is particularly powerful for this purpose [15-17]. Wang-Landau algorithm has been implemented in various type of systems, such Ising model, AF Potts model and fully frustrated Clock model [9-17].

Below we give the description of the Wang-Landau MC algorithm as implemented in the present work. In WL algorithm a random walk in energy space is performed with a probability proportional to the reciprocal of the DOS, 1/g(E), which results in a flat histogram of energy distribution. Actually, we make a move based on the transition probability from energy level $E_1$ to $E_2$:

$$p(E_1 \to E_2) = \min \begin{cases} g(E_1)/g(E_2) \\ 1 \end{cases}, \qquad (2)$$

Since the exact form of g(E) is not known a priori, we determine g(E) iteratively. We start g(E) = 1 and energy histogram H(E) = 0. Introducing the modification factor fi, g(E) is modified by

$$g(E) \to g(E) + \ln f_i, \qquad (3)$$

Every time the state E is visited. At the same time the energy histogram H(E) is updated as

$$H(E) \to H(E) + 1. \qquad (4)$$

The modification factor fi initialize by start value $f_0 = e$ and gradually reduced to unity by checking the "flatness" of the energy histogram. The "flatness" is checked such that the histogram for all E is not less than some value of the average histogram, e.g., 80%. Then $f_i$ is modified as

$$\ln f_{i+1} = \frac{1}{2} \ln f_i, \qquad (5)$$

and the histogram H(E) reset. A final value we choose $\ln f_i = 2^{-24}$.

After modification step i = 15 we analyze the ordering in energy minima and save it magnetic structure. A more detailed description for the Wang-Landau algorithm is reported in work [9, 10, 15].

## 3. Simulation Results

Thus, using the data for density of states we can calculate the thermodynamic parameter values at any temperature. Particularly, the internal energy $U$, the free energy $F$, entropy $S$ and the heat capacity $C$ can be calculated by equations

$$U(T) = \frac{\sum_E E g(E) e^{-E/k_B T}}{\sum_E g(E) e^{-E/k_B T}} \equiv <E>_T, \qquad (6)$$

$$F(T) = -k_B T \ln \left( \sum_E g(E) e^{-E/k_B T} \right), \qquad (7)$$

$$S(T) = \frac{U(T) - F(T)}{T}, \qquad (8)$$

$$C = (N\beta^2)\left(\langle E^2 \rangle - \langle E \rangle^2\right), \qquad (9)$$

where $\beta = |J_1|/k_B T$, N is the number of sites.

Calculations have been carried out for systems with periodic boundary conditions and linear size LxL = N, where L = 12, 24, 36 and 72. In this paper most data show for systems with linear size L = 36.

The ground state (minimal possible) energy of system on various $J_2$ show Fig. 2. As follow from fig 2, there are three phases for $J_1 = -1$ and $J_1 = 1$. For negative $J_1$ we have Phase_1 in $-\infty \leq J_2 < -1.0$, fully frustrated phase in $-1 \leq J_2 \leq -0.2$, and antiferromagnetic phase in $-0.2 < J_2 \leq \infty$. For positive $J_1$ we have Phase_1 in $-\infty \leq J_2 < -1.0$, frustrated phase in $-1 \leq J_2 \leq -0.5$ with fully frustrated points $J_2 = -1$, and ferromagnetic phase in $-0.5 < J_2 \leq \infty$. The magnetic structure of the all phase analyzed below.

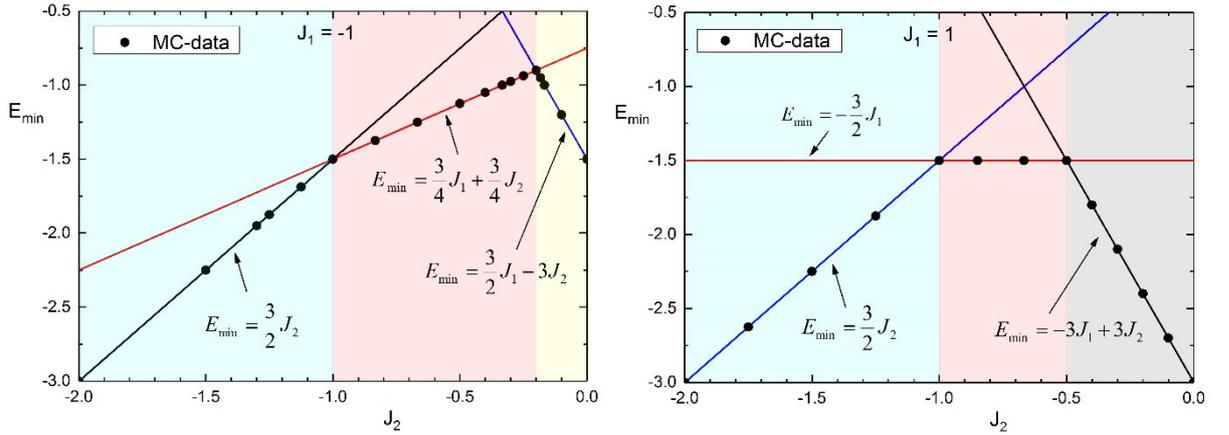

Fig. 2. Ground state (minimal possible) energies as function of $J_2$.

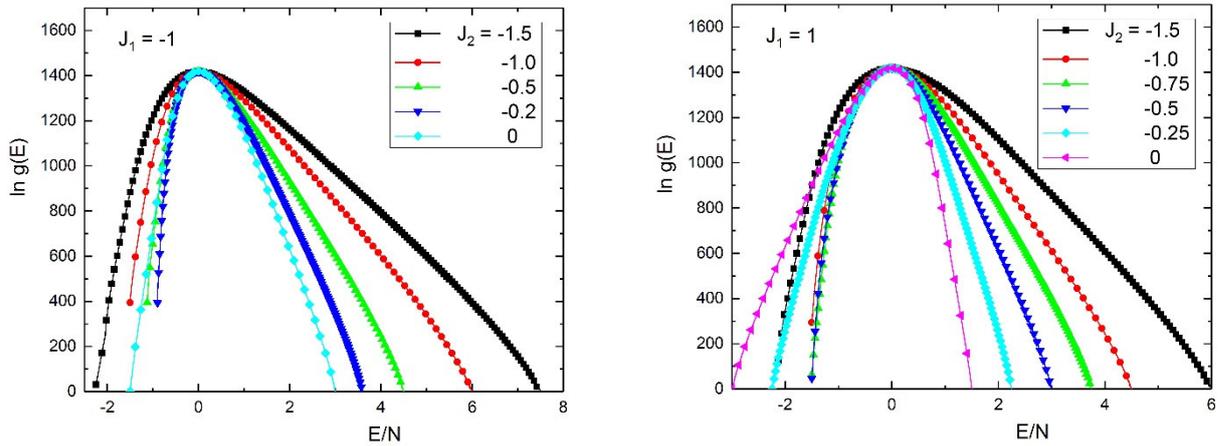

Fig. 3. Density of states for various $J_1$ and $J_2$.

The density of states for systems is shown in Fig. 3 (hereafter a statistic error doesn`t exceed symbol sizes used for dependences plotting). As a matter of convenience only a few of data are symbolized, all residue data fall on the line. The diagram depicts that for negative $J_1$ the high degeneration of the ground state are in area $-1 \leq J_2 \leq -0.2$ where $\ln g(E_{\min})$ stay by equal 400. This region are fully frustrated. On the right panel density of states come large value $\ln g(E_{min}) \approx 300$ only for $J_2 = -1$.

We calculate $\ln g(E)$ with WL algorithm and consider the difference of $\ln g(E)$, which is defined as [18]

$$\Delta \ln g(E)/\Delta E = [\ln g(E + \Delta E) - \ln g(E)]/\Delta E \qquad (10)$$

The difference of $\ln g(E)$ allow to justify the order of phase transition and determine the temperature of phase transition. We calculate DOSD for $\Delta E = 10$, $\Delta E = 1$ and $\Delta E = 0.1$. For large $\Delta E$ the DOSD are smooth, for small $\Delta E$ DOSD becomes the differential. The systems that show the first-order transition has an S-like structure of DOSD. The transition temperature $T_C = 1/\beta_C$, can be estimated by Maxwell's rule as in thermodynamics. Line on $\beta_C$ separate the DOSD on two zone with same area, as show in fig. 4. The value of critical temperature $T_C = 0.94648$, determined from Fig 4 are high accuracy.

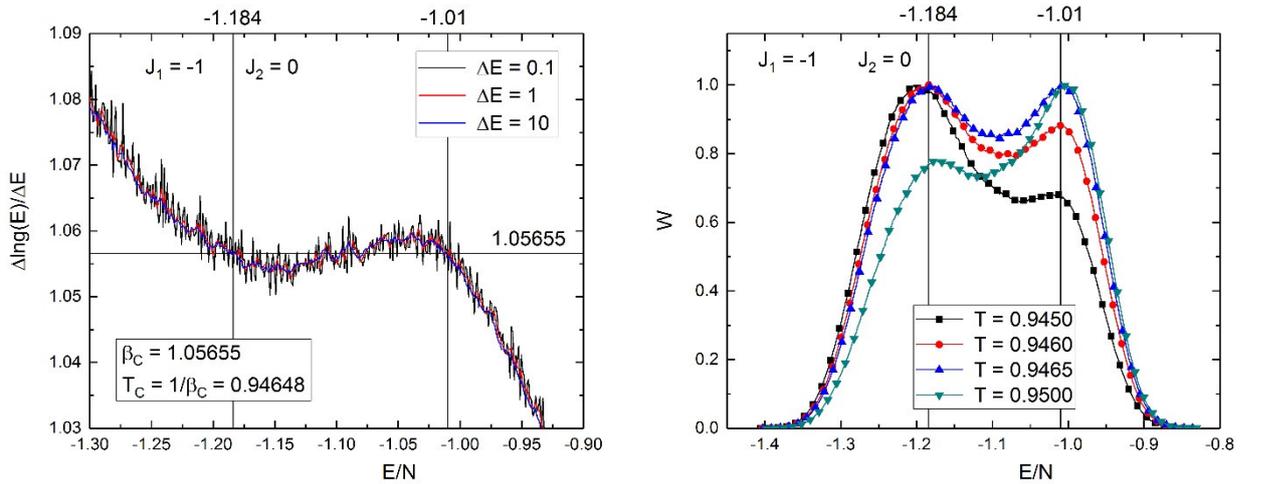

Fig. 4. Density of states difference $\Delta \ln g(E)/\Delta E$ (left panel) and the energy histogram W (right panel).

Also for determine order of the phase transition we use the histogram analysis [19-20]. The systems that show the first-order transition has a double-peak structure of histogram on critical temperature $T_C$.

The histogram of energy calculated as

$$W = g(E)e^{-E/k_B T}, \qquad (11)$$

and normalized to unity.

Energy histogram W on left panel of the Fig 4 have duble-peak structure for temperature T = 0.9465. This maximum come to energy $E_1$ = -1.184 and $E_2$ = -1.010 with good agrement of data from DOSD on left panel. The same procedure are carried out for all value $J_2$ and critical temperature and order of phase transtition calculated. Also for determine critical temperature we use the maxumum of the heat capasity.

Temperature dependence of the entropy S in frustrated region for negative $J_1$ (left panel) and positive $J_2$ (right panel). For negative $J_1$ and $-1 \leq J_2 \leq -0.2$ we have zero temperature entropy $S_0 \approx 0.3042$. For positive $J_1$ are only one points $J_2 = -1$ with arbitrary large zero temperature entropy $S_0 \approx 0.2267$. The entropy tends to predicted value ln3 at increasing the temperature.

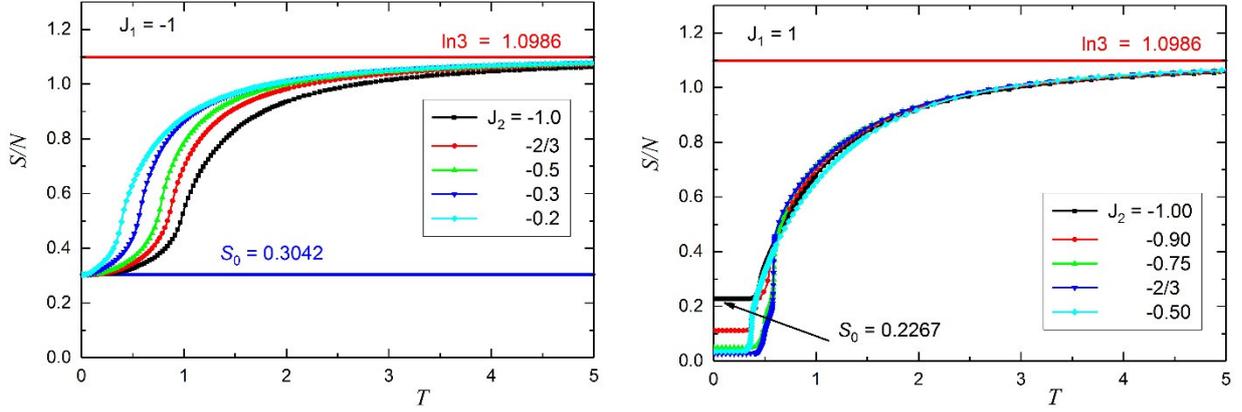

Fig. 5. Temperature dependence of the entropy S in frustrated region for various $J_1$ and $J_2$.

Ground state magnetic structure of the q=3 Potts model on triangular lattice show on Fig 6. For various $J_1$ and $J_2$ there are 8 structure:

**a)** Striped structure; **b)** Triplet structure; **c)** Striped structure; **d)** Fully frustrated nonordered structure; **e)** Partially frustrated nonordered structure; **f)** Partially frustrated nonordered structure;

**g)** Ferromagnetic ordered structure; **h)** Antiferromagnetic ordered structure.

Structure a), b) and c) have same energy $E_{min} = 0 J_1 + \frac{3}{2} J_2$ for both positive and negative $J_1$ and realized if $J_2 < -1$ for both negative and positive $J_1$. Structure d) with energy $E_{min} = \frac{3}{4} J_1 + \frac{3}{2} J_2$ realized in fully frustrated area $[J_1 = -1, -1 \leq J_2 \leq -0.2]$ as well as in point $[J_1 = 1, J_2 = -1]$. Structure e) with energy $E_{min} = \frac{3}{4} J_1 + \frac{3}{2} J_2$ realized in partially frustrated area $[J_1 = 1, -1 < J_2 < -0.5]$. Structure f) with energy $E_{min} = \frac{3}{4} J_1 + \frac{3}{2} J_2$ realized in partially frustrated point $[J_1 = 1, J_2 = -0.5]$. Structure g) correspond to ordered ferromagnetic phase with energy $E_{min} = -3 J_1 + 3 J_2$ realized in area $[J_1 = 1, J_2 > -0.5]$. Structure h) correspond to ordered antiferromagnetic phase with energy $E_{min} = \frac{3}{2} J_1 - 3 J_2$ and realized in area $[J_1 = 1, J_2 > -0.5]$.

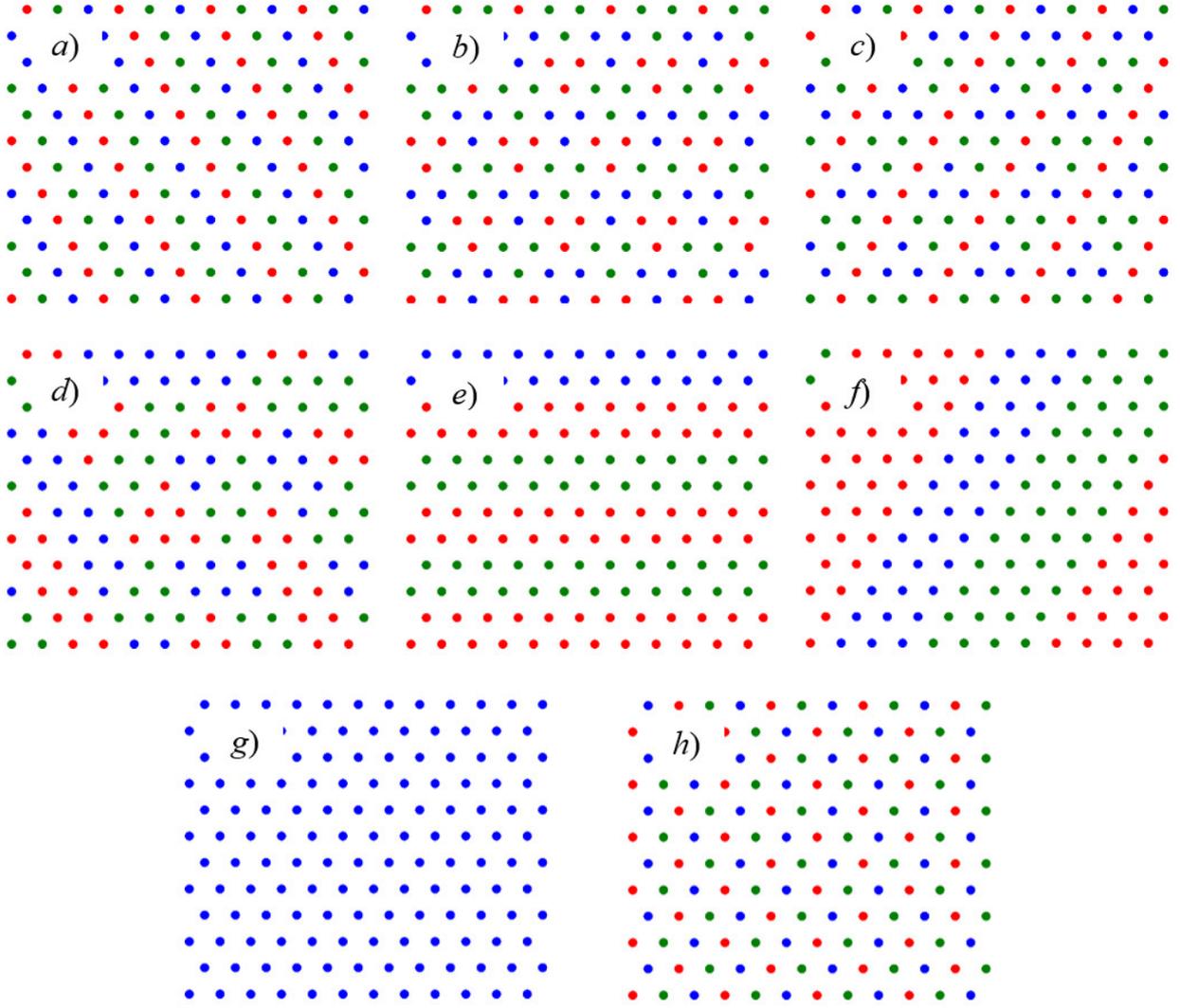

Fig. 6. The ground state magnetic structures.

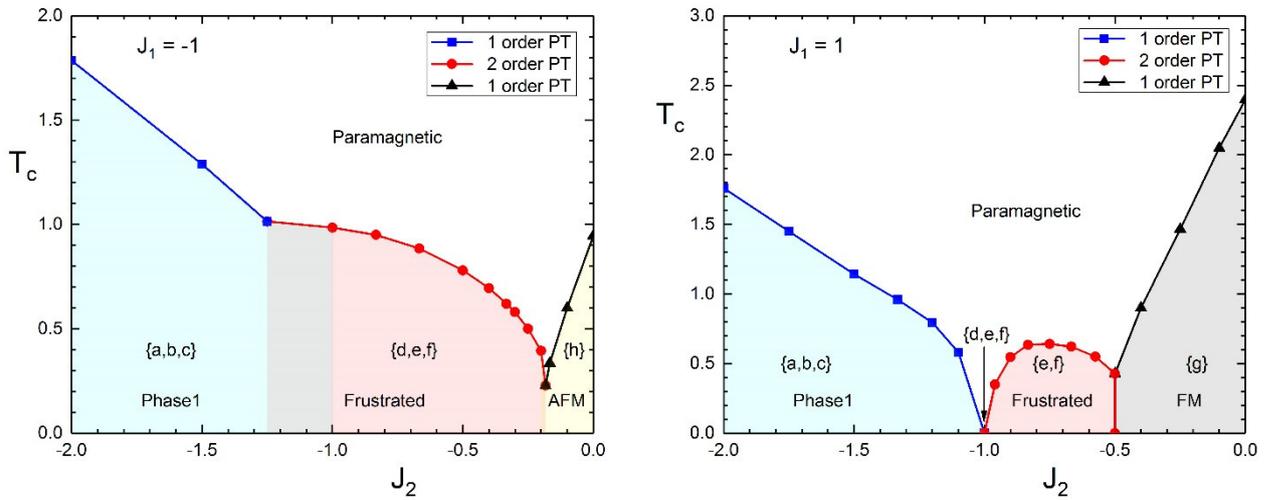

Fig. 7. The phase diagram of the three-state Potts model on triangular lattice.

Finally, on Figure 7 we show the phase diagram of the investigated system. On figure for each phase in bracket given corresponding ground state magnetic structure from figure 6. Gray colored area $[J_1 = -1, \; -1.25 \leq J_2 < -1]$ in left panel of Figure 7 have same a Phase1 ordered ground state structure, though closely located high degenerate frustrated energies levels, therefore the thermodynamic behaviors of system are such as frustrated, closely to near zero temperature. Order of phase transitions on Figure 7 are determined by difference of density of states and histogram analysis method (example are show on Figure 4).

## 4. Conclusion

We studied the three-state Potts model on triangular lattice with next-nearest-neighbor interactions using the high-performance Monte Carlo Wang-Landau algorithm. One main result summarized in figure 6 where show the ground state magnetic structure for various $J_1$ and $J_2$. We analyze all ordered structure in this model, and calculate ground state energies as well as zero temperature entropy. We determine the order of phase transition and calculate critical temperature using density of states difference and histogram analysis methods. The phase diagram are show.